\documentclass[twocolumn]{optica-article}

\journal{opticajournal} % for journals or Optica Open

\articletype{Research Article}

\usepackage{lineno}
\begin{document}

\title{A Reconstruction of Michelson Interference from On--Off Detection Statistics}

\author{Kim Myung-Whun\authormark{1,2*}}

\address{\authormark{1} Department of Physics, Jeonbuk National University, Jeonju 54896, Republic of Korea\\
\authormark{2} Institute of Photonics and Information Technology, Jeonju 54896, Republic of Korea}

\email{\authormark{*} mwkim@jbnu.ac.kr} %% email address is required; see note below about the corresponding author designation

% use {asbstract*} to suppress the copyright line. Copyright information will be added in production

\begin{abstract*} 
This paper reconstructs the Michelson interference phenomenon based solely on the binary (on-off) statistics of a single-photon counter and macroscopic path-difference data, without the a priori assumptions of complex wave superposition or prior spectral information. In contrast to the conventional approach that assumes light as a continuous complex wave function and considers interference between them, this study focuses on the algebraic properties inherent in discrete detection statistics. We verify that the discrete response data collected across various path differences form a positive semi-definite matrix, and apply Herglotz's theorem to this result. Consequently, we show that the harmonic form of the interference fringe is not a consequence of adopting a wave model as an axiom, but rather it is uniquely decomposed into a cosine average of wavenumbers dictated by testable algebraic constraints satisfied by the statistical data. Through this approach, the logical basis of the interference phenomenon can be described by the characteristics of the binary data itself, perfectly aligning with the modern optical experimental environment that relies entirely on discrete event statistics.
\end{abstract*}

%%%%%%%%%%%%%%%%%%%%%%%%%%  body  %%%%%%%%%%%%%%%%%%%%%%%%%%
\section{Introduction}
The Michelson interferometer measures properties of light through the changes in fringes produced by manipulating the path difference. The standard account of this phenomenon, in both classical and quantum optics, treats light as a complex wave and explains that the superposition of the amplitudes traversing the two paths, their phase difference, and the square of the modulus of their sum determine the final intensity. The complex amplitude, a continuous mathematical instrument, was an efficient device for describing the phenomenon in an era when intensity was measured with analogue photometers. Modern optics, however, has been reduced by the development of single-photon counting to a statistics of discrete events rather than a continuous flow of energy. In today's laboratories the primitive datum of measurement is the binary statistics of the presence ($1$) and absence ($0$) of a bit emitted by devices such as single-photon avalanche diodes. It is therefore worth attempting to reconstruct the logical basis of interference directly from the control variable, the displacement of the mirror, and the raw measured outcome, the binary statistics, rather than converting the discrete and direct sequence of binary signals into the continuous function that is the complex amplitude.

The present approach, based on ensembles of binary detections and a control variable, is in line with the operationalist tradition that seeks to specify the objects of a theory through measurement statistics alone~\cite{Ludwig1983} and with the epistemic view that reads the quantum state as a description of information rather than as a physical entity~\cite{Spekkens2007,Chiribella2011}. It differs, however, from event-based algorithmic models that reproduce the observed statistics by assigning particles an internal phase inaccessible to measurement or by treating detectors as adaptive devices~\cite{DeRaedt2005}. Rather than introducing a hidden mechanism or an a priori model, we examine the algebraic properties satisfied by the acquired binary statistics and identify the mathematical form of the response function that necessarily follows from them. In the standard account a single instrument, the complex amplitude, serves at once for the description of the state, for propagation, and for the derivation of probabilities, so that it is difficult to separate the conditions genuinely required for the physics from the devices introduced for computational convenience. In the present approach the operating conditions confirmed by experiment, the calibration values obtained from prior measurement, and the algebraic constraint on the response matrix built from binary statistics are kept distinct, which makes it easier to identify what within the structure of the account is subject to empirical refutation.

This paper derives interference from the algebraic properties of measured binary statistics alone, in place of the analogue instrument of complex wave superposition. We first define the physical measuring conditions of the apparatus and the calibration values, and separate the observed firing probability into a baseline and a response function. We then verify empirically that the discrete response statistics acquired at several path differences form a positive semi-definite matrix. Applying Herglotz's theorem to the positive semi-definiteness of this discrete matrix, we show that the response data decompose naturally into harmonic components without any separate assumption of continuity. Through this route the concepts of wavelength and of harmonic function are shown to be necessary consequences of a testable algebraic property, and we obtain an operational account grounded in statistical facts and suited to a digital measuring environment. 

\section{Apparatus and Measuring Conditions}
\label{sec:apparatus}

\subsection{Layout of the Apparatus}

\begin{figure}[htbp]
    \centering
    \includegraphics[width=0.7\linewidth]{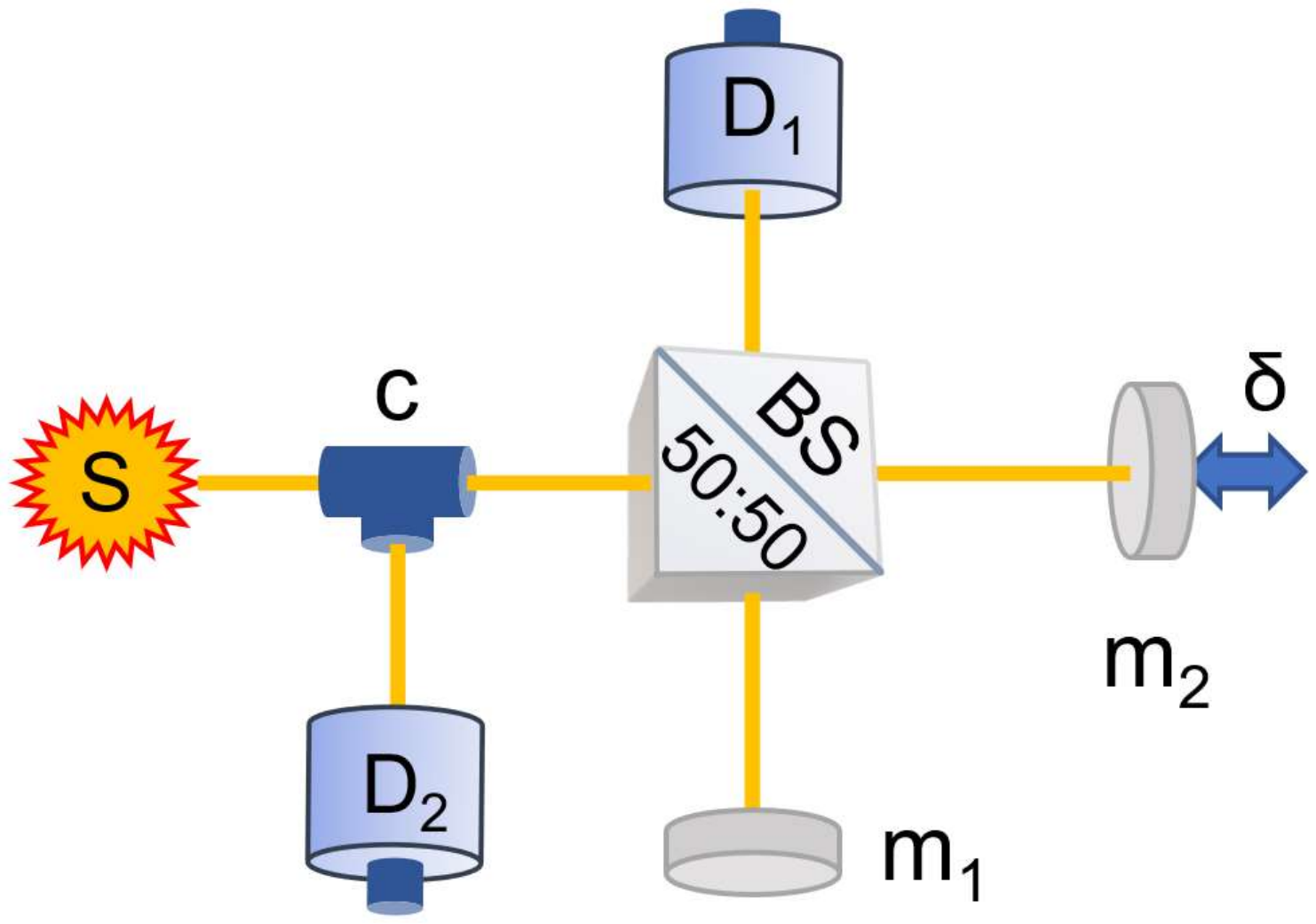}
    \caption{Michelson interferometer equipped with binary detectors. S is the source, C the optical circulator, BS the beam splitter, $m_1$ the fixed mirror and $m_2$ the movable mirror. $D_1$ is the detector at the output orthogonal to the direction of incidence and $D_2$ the detector at the output returning towards the source. The path difference $\delta$ is twice the displacement of $m_2$ and is read from a ruler.}
    \label{fig:1}
\end{figure}

Figure~\ref{fig:1} shows the layout of a conventional Michelson interference experiment. The signal leaving the source S passes through the optical circulator C, reaches the beam splitter BS, and divides into two branches. One branch is reflected at BS, reaches the fixed mirror $m_1$, returns, and is transmitted through BS towards the detector $D_1$; we call this arm 1. The other branch is transmitted through BS, reaches the movable mirror $m_2$, returns, and is reflected at BS towards the same detector $D_1$; we call this arm 2. Signals that have travelled two different paths thus meet at a single detector. Part of what meets at BS returns towards the source and enters the circulator, which admits the source while diverting to the detector $D_2$ only the signal returning from the direction of the beam splitter. It is equally possible to tilt the angle of incidence slightly and separate the returning beam spatially towards $D_2$; the discussion below applies to either arrangement, and for convenience we consider the case in which a circulator is used. When the mirror $m_2$ is displaced, the relative lengths of the two branches change. Writing $x$ for the displacement, the signal makes a round trip, so the path difference is $\delta = 2x$, and this value is read from a ruler. It is the only quantity the experimenter controls. This quantity is sometimes called a time delay, but what is actually controlled and acquired is a macroscopic length.

The detector considered in this work is a binary device that resolves neither the energy of the light that reaches it nor the number of photons. The element fires and emits $1$ when given a stimulus above its threshold, and remains silent and emits $0$ otherwise. The inability to resolve photon number may be regarded as a limitation of the instrument, but as noted in the introduction it works to our advantage here: because the data are binary, the properties needed to treat probabilities are secured without any separate assumption.

To handle binary events statistically on a probability space, the continuously flowing time axis must be divided into a countable ensemble of trials. The width of the time bin is fixed in advance by the apparatus, independently of when and how often photons arrive. Real detectors and counting circuits have their own timing jitter and dead time, and these limitations impose a minimum time window, the time bin $\Delta t$, that partitions the record. Cutting the total observation time into a discrete sequence of such windows makes one time bin one individual trial, whose outcome is either firing or not firing. The signal a detector actually emits is not a mathematical point but a voltage pulse with a rise, a plateau and a fall, spread over a finite time; converting it into a single instant requires a discriminator circuit, and jitter arises in that process. The present account does not address errors due to jitter, because it uses not the time of arrival but only whether firing occurred in each time bin. There are cases in which a pulse straddles a bin boundary so that its assignment is ambiguous, but such ambiguity is independent of the path difference and therefore does not bias the response function defined below. Some bins receive a photon and others do not, and time bins in which nothing happened are included among the trials that constitute the whole data ensemble. What this experiment leaves is not a continuous quantity of light but an array of $1$s and $0$s recorded over a sequence of time bins. At each path difference the frequency with which the detector fired is the ratio of the number of firings, that is, of bins recording $1$, to the total number of time bins, and we call this value the firing probability of the detector.

\subsection{Measuring Conditions}
\label{sec:conditions}

We assume that the experiment is performed in a restricted environment in which the following six conditions are satisfied. These are conditions to be respected when the apparatus is assembled and while data are acquired, and all of them are confirmed by measurement.

\begin{enumerate}
\item Photons do not arrive from the source in bunches. Using a parametric down-conversion source at extremely low intensity places one in this regime.
\item The experiment is performed in a regime in which the firing probability of the detector is sufficiently small. As explained again below, this condition assumes that, even with an ideal detector, photons are incident so rarely that the majority of bins contain no firing signal; this is not an unusual condition.
\item The two end mirrors have equal reflectivities and the optical losses of the two arms are equal. An imbalance in the splitting ratio of the beam splitter cancels structurally, but the effects of imperfect optical elements or of a medium placed asymmetrically along a path do not. This is confirmed by comparing the firing frequencies measured with each arm blocked in turn.
\item Displacing the mirror does not alter the statistics of the source or the loss rate of the apparatus. This is confirmed by repeating the blocking measurement at different positions of the movable mirror and checking that the counts remain constant.
\item The beam splitter does not absorb, so that its transmittance $T$ and reflectance $R$ sum to unity. The beam splitter also does not split a photon: a photon entering during one time bin causes a firing at only one of the two outputs. Whether this condition holds is checked by confirming that the two detectors never fire in the same time bin, although this requires a separate experiment.
\item The dark count of the detectors is sufficiently small. A large dark count produces a background signal unrelated to the path difference, and we exclude such cases. Since the dark count is a constant component that does not respond to the path difference, it makes the magnitude of the firing probability appear smaller than it is. It is measured directly by blocking both arms and counting the firing frequency, and can therefore be corrected for.
\end{enumerate}

Apart from the character of the detectors, these conditions are no different from those of a conventional Michelson interference experiment.

\subsection{Calibration Measurement}
\label{sec:calib}

We assume that a calibration measurement is carried out separately from the interference experiment in order to establish the characteristics of the apparatus and the conditions of the experiment. The calibration consists in blocking each arm in turn and reading the counts of both detectors. Blocking arm 2 leaves only the branch that is reflected at BS, travels to $m_1$ and returns. We denote the firing frequency of $D_1$ in this configuration by $w_1$ and that of $D_2$ by $b_1$. Blocking arm 1 and performing the same measurement yields $w_2$ and $b_2$. In both measurements only one path is open, so there is no room for the path difference to enter and the values are independent of $\delta$. These four numbers are the only prior measurements used in this paper. The absolute brightness and spectral density of the source, the scattering losses of the optics, the quantum efficiency of the detectors and the efficiency of the circulator are all absorbed and condensed into these values, and we do not track them separately.

In the conventional interferometric layout of Fig.~\ref{fig:1} the weights of the two paths are balanced. The way the two branches traverse their paths shows why $w_1$ and $w_2$ are balanced. The first branch, associated with $w_1$, is reflected on entry and transmitted on return, while the second branch, associated with $w_2$, is transmitted and then reflected. Both branches undergo one transmission and one reflection, so both weights are proportional to $TR$. This balance is maintained even when the beam splitter departs from symmetry so that $T \ne R$. It is not achieved by adjusting the apparatus but is guaranteed by the geometry of the layout. Because the two branches are symmetric, there is no room to infer which path was taken from any imbalance in the counts. To simplify matters we consider here the ideal symmetric arrangement in which the balance of the two arms is exact, and we write $w_1 = w_2 \equiv w$ and $b_1 = b_2 \equiv b$. We also take the transmittance and reflectance of the beam splitter to be symmetric, $T = R = 1/2$; with an ideal optical circulator this implies $w = b$.

Measuring the firing probabilities $P_1(\delta)$ and $P_2(\delta)$ of $D_1$ and $D_2$ together allows a consistency test on the probabilities. The test rests on the fact that the apparatus does not absorb photons along the way. A photon leaving the beam splitter on its return exits at only one of the two outputs, so by the conditions above it is lost nowhere. The sum of the firing probabilities at the two outputs must therefore equal the probability that a photon entered the beam splitter, and this value is independent of the position of the mirror, so that a summation relation independent of the path difference must hold. Let $p$ be the probability of entry into the beam splitter per time bin. Blocking arm 2 leaves only the branch that is reflected at the beam splitter, travels to $m_1$ and returns. If that branch is transmitted when it meets the beam splitter again it goes to $D_1$ with probability $pRT$, so that $w = p/4$; if it is reflected it goes towards $D_2$ with probability $pRR$, so that $b = p/4$. In the symmetric arrangement, blocking arm 1 likewise sends probability $w$ to $D_1$ and $b$ to $D_2$. Adding the probabilities gives

\begin{equation}
P_1(\delta) + P_2(\delta) = 4w = p
\label{eq:sumrule}
\end{equation}

\noindent at every path difference. This relation can be verified experimentally, and if it fails the apparatus is losing photons somewhere, which we regard as a situation in which the present account does not apply.

Among the measuring conditions stated above, we take $4w \ll 1$ as the criterion for the firing probability being sufficiently small. Our account assumes that the firing probability of the detector is proportional to the number of photons reaching it, but a real threshold detector behaves linearly only when certain conditions are met. Assuming the photons follow Poisson statistics, if the mean number of photons actually reaching the detector per unit time with a single path open is $\bar n$, the probability that the detector fires is $w = 1 - e^{-\bar n}$. If the number of photons entering the detector is quadrupled to $4\bar n$, the firing probability does not increase fourfold but is predicted to be $1 - e^{-4\bar n}$, which is not linear. If the firing probability is very large, say $w = 0.2$, the relative error between the linear assumption and the actual prediction, $\bigl|4w-(1-e^{-4\bar n})\bigr|/4w$, is about $26\%$, and our assumption fits poorly. For $w = 0.01$, however, the error is about $1.5\%$, and for $w = 0.001$ it is $0.15\%$. In other words, for the simple additive account of Eq.~\eqref{eq:sumrule} to reach an accuracy at the level of one per cent or better, the firing probability must be held below $0.01$ so that $w$ and $\bar n$ nearly coincide, and the counting time must be lengthened accordingly. Given that a time bin is about $10$~ns under ordinary conditions, this corresponds to roughly one photon reaching the detector per microsecond. Such a condition is not unusual and is standard in single-photon counting experiments~\cite{Achilles2003}. Whether the condition is met can be checked by varying the source intensity and seeing whether the firing probability is proportional to it.

\subsection{Definition of the Wavenumber}

If the wavenumber band of the source is narrow, or is made narrow with a narrowband filter, the firing probability $P_1(\delta)$ oscillates with a nearly constant period. In such conditions, writing $N_{\rm f}$ for the number of fringes counted while the path difference is changed by $\Delta$, the wavenumber $\kappa$ can be defined from a length read on a ruler and a number of fringes read on a counter,

\begin{equation}
    \kappa \equiv \frac{2\pi N_{\rm f}}{\Delta}.
    \label{eq:kappa}
\end{equation}

\noindent In what follows we use the wavenumber as a value defined in this way through the measurement.

\section{Prediction of the Measured Results}
\label{sec:response}

\subsection{Effect of the Path Difference}

As the movable mirror is displaced to increase the path difference, one observes, for a source containing light of several colours, that the fringes gradually subside and the firing probability converges to a constant value. The value to which it converges equals the simple sum of the results of the blocking measurements. How rapidly the fringes subside depends on the source; the closer the source is to monochromatic, the further the fringes survive in path difference. Thus

\begin{equation}
P_1 = w_1 + w_2 = 2w
\label{eq:distinct}
\end{equation}

\noindent is the baseline of $P_1$.

Conversely, as the path difference is reduced, an oscillation is observed in the firing probability. To describe the observed path-dependent variation we separate the firing probability into the baseline $2w$ and a variation $f$ around it,

\begin{equation}
P_1(\delta) = 2w + f(\delta),
\label{eq:master0}
\end{equation}

\noindent without yet knowing the form of $f$. Its range, however, can be fixed from measured facts. Since a firing probability cannot be negative, $f(\delta) \ge -2w$; and since our apparatus loses no photons along the way, the summation relation of Eq.~\eqref{eq:sumrule} gives $P_1 \le 4w$ and hence $f(\delta) \le 2w$. Thus

\begin{equation}
-2w \le f(\delta) \le 2w
\label{eq:fbound}
\end{equation}

\noindent bounds the variation. Since the magnitude of the variation is confined to $2w$, it is natural to adopt that value as the unit. Setting

\begin{equation}
g(\delta) \equiv \frac{f(\delta)}{2w} = \frac{P_1(\delta)}{2w} - 1, \qquad -1 \le g(\delta) \le 1,
\label{eq:measure}
\end{equation}

\noindent the firing probability takes the form

\begin{equation}
P_1(\delta) = 2w\,\bigl[1 + g(\delta)\bigr].
\label{eq:master}
\end{equation}

\noindent The response function $g(\delta)$ is a quantity measured directly at the detector, and its functional form is fixed by the properties of the source.

The values $g(\delta)$ are measurements obtained one at a time for each path difference. Let us collect these values in a suitable way and look for an algebraic relation among them. First, lengthening arm 1 ($\delta > 0$) and shortening it ($\delta < 0$) are geometrically symmetric with respect to a displacement of arm 2. The measured response function must therefore be even under sign reversal, $g(-\delta) = g(\delta)$. If the mirror is pushed in one direction in constant steps of $h/2$ while the response function is measured, then, since the signal makes a round trip in the arm, one obtains the values $g(0), g(h), g(2h), \dots$ at path differences $\delta_k = k \cdot h$ with $k = 0, 1, 2, 3, \dots$. From these data one can build the symmetric matrix whose elements are $g\bigl((i-j)h\bigr)$,

\begin{equation}
    C = \begin{pmatrix}
     g(0)  & g(h)  & g(2h) & g(3h) & \cdots \\
     g(h)  & g(0)  & g(h)  & g(2h) & \cdots \\
     g(2h) & g(h)  & g(0)  & g(h)  & \cdots \\
     g(3h) & g(2h) & g(h)  & g(0)  & \cdots \\
     \vdots & \vdots & \vdots & \vdots & \ddots
    \end{pmatrix}.
\end{equation}

As Eq.~\eqref{eq:measure} shows, this is a symmetric matrix of real numbers between $-1$ and $1$. If none of its eigenvalues is negative, the matrix is positive semi-definite. The elements of the matrix consist of the values of some response function $g(\delta)$, but an arbitrary function of $\delta$ does not make $C$ positive semi-definite. If, for instance, $g(\delta)$ has a triangular or rectangular shape, negative eigenvalues appear among those of the matrix built from its values. Positive semi-definiteness is a property that can be checked directly from the measured values, and this condition alone restricts the mathematical forms the response function may take.

If the matrix $C$ is positive semi-definite, then so is every one of its $2\times2$ minors. Taking the minor formed from the two points $0$ and $\delta$, whose diagonal elements are both $g(0)$ and whose off-diagonal elements are both $g(\delta)$, the requirement that its determinant be non-negative gives

\begin{equation}
g(0)^2 - g(\delta)^2 \ge 0 \;\Longrightarrow\; |g(\delta)| \le g(0).
\label{eq:maxatzero}
\end{equation}

\noindent The response function therefore takes its largest value where there is no path difference and exceeds that value at no other $\delta$. Positive semi-definiteness tells us only that the values at other $\delta$ cannot exceed $g(0)$; the value of $g(0)$ itself must be fixed by measurement.

\subsection{Relation between the Response Function and the Source}
\label{sec:herglotz}

In a conventional Michelson interferometer the oscillation of $P_1(\delta)$ takes a rather complicated form, and what determines that form is the response function $g(\delta)$. When the values of $g(\delta)$ measured at several path differences satisfy the positive semi-definiteness condition, Herglotz's theorem determines this response curve uniquely as an average of cosines of several wavenumbers with definite weights~\cite{Herglotz1911,Bochner1933}. To apply the theorem the sequence must be real and even and its first term must be unity. The measured response function $g(\delta)$ is real, since it is a ratio of firing frequencies, and it is even, $g(-\delta) = g(\delta)$, since by the symmetry of the two arms it does not depend on the direction in which the mirror is displaced. The value $g(0)$ is determined by measurement, and in order to set it to unity we introduce the normalised function

\begin{equation}
    \hat g(\delta) \equiv \frac{g(\delta)}{g(0)}.
    \label{eq:normalized}
\end{equation}

Applying Herglotz's theorem to this normalised $\hat g(\delta)$ yields the functional form of $g(\delta)$. In an ordinary measurement a spectrometer divides wavenumber space into a finite number of pixels and obtains a count histogram at each pixel, so the measured data are discrete and no values are assumed for the intervals that were not measured. The wavenumber is measured according to Eq.~\eqref{eq:kappa}, and since the number of measured values is necessarily finite, the number of terms in the response function constructed from them is finite as well. Writing $\mu_1, \dots, \mu_M$ for the weights assigned to the wavenumbers $\kappa_1, \dots, \kappa_M$, Herglotz's theorem allows the following discrete expansion using positive wavenumbers alone,

\begin{equation}
    \hat g(\delta) = \sum_{i=1}^{M} \mu_i \cos(\kappa_i \delta).
    \label{eq:herglotzexp}
\end{equation}

\noindent By the theorem the weights $\mu_i$ cannot be negative, $\mu_i \ge 0$, and they sum to unity, $\sum \mu_i = 1$. This collection of non-negative shares forms the wavenumber distribution of the source. The theorem confines the interval on which the wavenumber distribution lives to $[-\pi/h, \pi/h]$, and since the negative interval can be excluded using evenness, the $\kappa_i$ of Eq.~\eqref{eq:herglotzexp} lie in $[0, \pi/h]$. If the wavenumbers of the source fall outside this range, different wavenumbers yield identical data and the distribution is not uniquely determined. The path-difference step $h$ must therefore be smaller than half the shortest wavelength contained in the source, and the actual displacement step of the mirror smaller than a quarter of that wavelength. This is the same condition as the sampling limit of Fourier transform spectroscopy, and it is a necessary consequence of the theorem. The theorem guarantees that the weights reproducing the observed values exist uniquely, and on the strength of that uniqueness the wavenumber distribution of the source can be inverted determinately from the response curve.

\subsection{Prediction of the Firing-Probability Curve}
\label{sec:predict}

Inverting Eq.~\eqref{eq:herglotzexp} with respect to the measured values yields the weights $\mu_i$, which amounts to using the interferometer as a spectroscopic instrument. Conversely, if the distribution of detection probability over wavenumber for the source, or for a spectral filter, has been measured beforehand with a separate spectrometer, the firing-probability curve can be predicted in the reverse direction. To distinguish them from the $\mu_i$ produced by the interferometer, we write $\nu_i$ for the fraction obtained in the $i$th interval of the spectrometer. Acquiring the reference firing probability $w$ from the blocking measurement and $g(0)$ from a single measurement at zero path difference, the firing probability is computed as

\begin{equation}
P_1(\delta) = 2w\Bigl[1 + g(0)\sum_{i=1}^{M} \nu_i \cos(\kappa_i \delta)\Bigr].
\label{eq:predict}
\end{equation}

\noindent This result agrees with what the standard accounts of classical optics and quantum mechanics predict. Here $\nu_i$ is the ratio obtained by dividing frequency space into a finite number $M$ of intervals, counting the photon-counting clicks in each interval and dividing by the total number of clicks, and $\kappa_i$ is the central wavenumber of that interval. These values are obtained directly from the count histogram of the spectrometer. What is predicted here is the shape of the curve; the overall magnitude, fixed by $g(0)$, can be matched by measuring a single point with the interferometer. The period of the oscillation, the width over which it subsides, and the shape produced by the mixing of several wavenumbers are all fixed by the $\nu_i$, so that once a single point has been matched the whole of the remainder is subject to test. The spectrometer measurement uses the same source as the interferometer, and the statistical condition that the firing probability in each wavenumber interval be sufficiently low must likewise be respected. The detector of the spectrometer may be the same binary photon counter used in the interferometer.

The key question that remains is whether the $\mu_i$ derived from the interferometer and the $\nu_i$ measured by the spectrometer agree. The two instruments operate on different principles. The measuring intervals of the spectrometer are fixed by its geometry, such as the spacing of the grating or its rotation angle, whereas the wavenumbers of the interferometer are fixed by the measured data, such as the displacement step of the mirror. Since the two sets of data are discretised according to different criteria, comparing the two results directly requires first reconstructing them on a common set of intervals. Moreover, the interferometer creates a path difference by displacing a mirror while the spectrometer uses diffraction or refraction, so the agreement of the two results is not self-evident. If, for example, there is a systematic error in the scale of the movable mirror, the position of the peak of the distribution inverted from the interferometer will depart from the value measured by the spectrometer, and this deviation indicates the magnitude of the scale error. Nevertheless, when a reliable spectrometer is used and extrapolation and apodisation are properly controlled in the data transformation, it is confirmed experimentally that the distributions obtained from the two instruments agree within their errors~\cite{mukai2021,Gattinger2025QFTIR}.

In sum, from the fact that the matrix $C$ assembled from the firing frequencies at several path differences satisfies the positive semi-definiteness condition, the response function decomposes into a mixture of harmonic components with weights. The collection of weights $\mu_i$ so derived, the wavenumber distribution, is an empirical quantity inverted from binary statistics. That the values $\nu_i$ measured by a spectrometer, based on an unrelated operation, agree with this distribution provides a cross-validation of the present framework and demonstrates that an operational reconstruction grounded in binary statistics and Herglotz's theorem is not a mere mathematical abstraction.

\section{Discussion}

On the basis of the binary statistics of a single-photon counter, this work has described two things: that the sum of the firing probabilities at the two outputs is independent of the path difference, and that the wavenumber distribution of the source determines the shape of the firing-probability curve. The former follows from the conservation of photons within the apparatus, the latter from the agreement between the measure supplied by the harmonic decomposition and the histogram of the spectrometer. Under ideal conditions such as perfect symmetry, every element of the framework, including the blocking measurement, is either an operating condition confirmed by experiment or a property checked by measurement. The wave function, the superposition of complex probability amplitudes and the assumption of continuous harmonic waves have been excluded. This is not to deny the existence of the wave function but to delimit how far an explanation can reach using operational facts and algebraic conditions alone. In the same spirit, rather than explaining why the response matrix is positive semi-definite, we have offered a method of checking it, through eigenvalue computation and cross-validation with a spectrometer, and derived its algebraic consequences. That standard wave mechanics confers this property automatically through the structure of the state space without asking after its origin places it on the same footing as the present approach; what differs is that the present framework states directly what must be checked empirically. All that this methodology guarantees is that a unique distribution of weights over wavenumber exists which reproduces the observed response curve.

In a real source there may be photon bunching, which we have not considered directly in the discussion. Bunching causes a binary threshold detector to register multiple photons as a single count, and can therefore lower the baseline below $2w$. This error, however, is controlled by the prior conditions of the framework. First, for both Poissonian and bunched sources the correction is of order $O(w^2)$, so the condition $4w \ll 1$ suppresses it. Second, the time bin of a typical apparatus is hundreds of times longer than the time scale on which photon bunching occurs in ordinary sources; even when several photons enter within one time bin, the counting statistics therefore converge to a Poisson distribution and the bunching effect cancels. A study of bunching effects is nonetheless worth pursuing in future work, and in particular it will be necessary to examine how the present approach must change when photon-number-resolving detectors are introduced. In that case the observable will be the mean count per time bin instead of the firing probability. By the linearity of expectation the contributions of the two paths combine by simple addition regardless of any correlation, so that the baseline correction and the condition $4w \ll 1$ become unnecessary. Even then the range of the response function is preserved, since that range follows from the non-negativity of the observable and from the summation relation of Eq.~\eqref{eq:sumrule}, and both grounds hold for mean counts as well. It would also be an interesting line of research not to leave the positive semi-definiteness of the response matrix as a merely empirical test, but to search for measuring conditions under which the property breaks down, or to explore the possibility of deriving it mathematically from more fundamental operational principles.

\section{Conclusion}

This work has presented a framework describing the interference of a Michelson interferometer using only the binary detection statistics of a photon counter and the path difference, which is the control variable. Assuming perfect symmetry, the baseline is established by independent prior measurement alone, and the remaining elements required by the account are all operating conditions checked by experiment. Within this framework the response function has a harmonic form, but that harmonic form is not derived from assuming the photon to be a complex wave. Once one confirms that the binary response statistics acquired at several path differences constitute a positive semi-definite matrix, it follows as an algebraic consequence that the observed response curve reduces uniquely to a weighted average of cosine functions over several wavenumbers. The collection of weights obtained by inverting the measured curve is the wavenumber distribution of the source, and the wavenumber is here specified as the empirical value obtained by dividing a length measured with a ruler by a number of fringes. This relation applies in both directions: inverting the wavenumber distribution from the binary count record of the interferometer, and predicting the firing-probability curve of the interferometer from a wavenumber distribution measured with a separate spectrometer. The agreement of these two results, based on the physically distinct operations of diffraction and path difference, constitutes an independent cross-validation of the consistency of the framework.

As the measurement of light shifts from continuous photometry to the statistics of discrete bit sequences, the system used to describe the phenomena should likewise conform to the primitive form of the empirical data. This work has shown that the harmonic form and the wavenumber distribution can be derived in full from the algebraic constraints that the binary statistics themselves impose, without the detour of converting the acquired bit sequence into continuous parameters or an a priori model. Whether this statistical and operational approach can be extended to more complex photon-counting experiments, and what mathematical changes the acquisition of photon-number resolution would bring to conclusions grounded in binary statistics, are left for future work. 

\begin{backmatter}
\bmsection{Funding}
This research was supported by the Jeonbuk National University Research Fund (2025).

\bmsection{Disclosures}
The authors declare no conflicts of interest.

\bmsection{Data Availability Statement}
No data were generated or analyzed in the presented research.
\end{backmatter}

\bibliography{bibliography}

\end{document}